\documentclass[
    ,final            
  ]
  {aipproc}

\layoutstyle{8x11double}

\begin{document}

\title{Thermodynamics of 5D dilaton-gravity}


\author{E.~Meg\'{\i}as}{
  address={Institute for Theoretical Physics, University of Heidelberg, Germany},altaddress={Instituto de F\'{\i}sica Te\'orica CSIC-UAM, Universidad Aut\'onoma de Madrid, Spain}
}

\begin{abstract}
We calculate the free energy, spatial string tension and Polyakov loop
of the gluon plasma using the dilaton potential of
Ref.~\cite{Galow:2009kw} in the dilaton-gravity theory of AdS/QCD. The
free energy is computed from the Black Hole solutions of the Einstein
equations in two ways: first, from the Bekenstein-Hawking
proportionality of the entropy with the area of the horizon, and
secondly from the Page-Hawking computation of the free energy. The
finite temperature behaviour of the spatial string tension and
Polyakov loop follow from the corresponding string theory in
$AdS_5$. Comparison with lattice data is made.
\end{abstract}

\maketitle


\section{Introduction}
\label{sec:introduction}

The duality of string theory with 5D gravity is nowadays a powerful
tool to study the strong coupling properties of gauges theories. In
conformal $AdS_5$ the metric has a horizon in the bulk space at $r_T=
\pi \ell^2 T$ where $\ell$ is the size of the AdS-space, and entropy
scale like $s \propto r_T^3 \propto T^3$. To extend this duality to
SU($N_c$) Yang-Mills theory, one of the first tasks is to control the
breaking of conformal invariance. In this work we will study the 5D
dilaton-gravity model introduced in Ref.~\cite{Gursoy:2008za}, with
the action
\begin{eqnarray}
{\cal S} &=& \frac{1}{16 \pi G_5}\int d^5x \sqrt{-G}\left(R-\frac{4}{3}\partial
_{\mu }\phi \partial ^{\mu }\phi -V(\phi )\right) \nonumber \\
&& -\frac{1}{8 \pi G_5}
\int _{\partial M} d^4x \sqrt{-H} K  \,. \label{eq:action5D}
\end{eqnarray}
The dilaton potential~$V(\phi)$ is related to the $\beta$-function in a one-to-one relation. We assume the parameterization
\begin{eqnarray}
  \beta(\alpha) &=& -b_2\alpha + \bigg[b_2\alpha  +
\left(\frac{b_2}{\bar{\alpha}}-\beta_0\right)\alpha^2 \nonumber \\
&&\quad +
\left(\frac{b_2}{2\bar{\alpha}^2} - \frac{\beta_0}{\bar{\alpha}}-\beta_1 \right)
\alpha^3 \bigg] e^{-\alpha/\bar{\alpha}}\,,
\label{eq:beta}
\end{eqnarray}
which was proposed in Ref.~\cite{Galow:2009kw} for the computation of
the $Q\bar{Q}$ potential at $T=0$ within this
model. Eq.~(\ref{eq:beta}) has the standard behavior in the UV,
$\beta(\alpha) \simeq -\beta_0 \alpha^2 - \beta_1 \alpha^3 + \dots$
for $\alpha \ll \bar{\alpha}$, and generates confinement in the
IR. The optimum values to reproduce the $Q\bar{Q}$ potential at $T=0$
are~$b_2 = 2.3$, $\bar\alpha = 0.45$ and $\ell = 4.389
\textrm{GeV}^{-1}$~\cite{Galow:2009kw}.

The model of Eq.~(\ref{eq:action5D}) has two types of solutions. The thermal gas solution corresponds to the confined phase, and the black hole (BH) solution characterizes the deconfined phase and it has a horizon in the bulk coordinate at $z=z_h$~\cite{Gursoy:2008za}. The BH metric takes the form
\begin{eqnarray}
ds^2       &=& b^2(z)\left(f(z) d\tau^2 + dx_kdx^k + \frac{dz^2}{f(z)}\right) \,, \label{eq:bhmetric}
\end{eqnarray}
where $z = \ell^2/r$ and $f(z)$ has the properties $f(z_h) = 0$ and
$f'(z_h) = -4\pi T$. Starting from the ultraviolet expansion of the
$\beta$-function, one can solve analytically the equations of motion
in the
UV~\cite{Gursoy:2008za,Alanen:2009xs,Megias:2010tj,Megias:2010ku}.
One arrives at the following expansion for the metric factor up to
${\cal O}(\alpha^2)$
\begin{equation}
b(\alpha) \stackrel{\alpha\to 0}{\simeq} \frac{\ell}{z} \left[  1 -\frac{4}{9} \beta_0 \alpha +\frac{2}{81}\left(22\beta_0^2 -9\beta_1 \right) \alpha^2  \right] \,. \label{eq:bzuv}
\end{equation}

\section{Free Energy}
\label{sec:FreeEnergy}

The phase transition from the confined glueball gas to the deconfined
gluon plasma can be understood as follows. At low temperatures the
physics is determined by the thermal gas metric with $f =1$. At a
minimal temperature the BH solution appears with a horizon at $z_h$
and $f \neq 1$, and becomes the preferred solution at the phase
transition. So, the phase transition arises from the competition of
the free energies computed in both metrics.

\begin{figure}[tbp]
\centerline{\includegraphics[width=7.3cm,height=5cm]{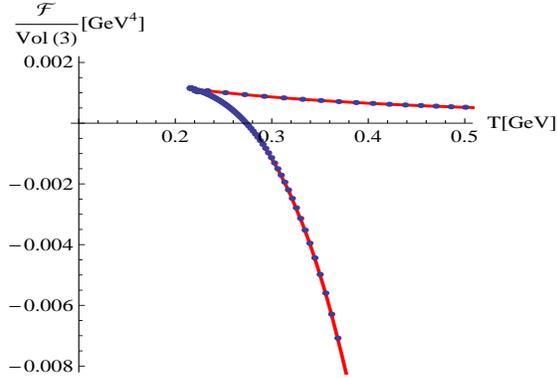}}
\caption{Free energy density as a function of temperature.  We show as a full
(red) line the numerical result obtained starting from the Bekenstein-Hawking
entropy formula, c.f. Eq.~(\ref{eq:s1text}). Blue points
correspond to the result using Eq.~(\ref{eq:Fk}). We include up to
${\cal O}(\alpha_0^2)$ in the r.h.s. of Eq.~(\ref{eq:bz4}) to compute $G$.}
\label{fig:freeenergy12}
\end{figure}

We show two different ways to get the thermodynamics of the 5D dilaton-gravity model of Eq.~(\ref{eq:action5D}). One possibility is to use the Bekenstein-Hawking entropy formula which establishes the proportionality between the entropy and the area of the horizon of the BH. It reads:
\begin{eqnarray}
&&s(T) = \frac{1}{4 G_5} b^3(z_h) \label{eq:s1text} \\
&\stackrel{\alpha_h \to 0}{\simeq}& \frac{\pi^3 \ell^3}{4 G_5} T^3
\Bigg[ 1 -\frac{4}{3} \beta_0 \alpha_h + \frac{1}{9} \left(
11\beta_0^2 -6\beta_1 \right) \alpha_h^2 \Bigg] \,, \nonumber
\end{eqnarray} 
where $\alpha_h = \alpha(z_h)$. In the second equality we have used
Eq.~(\ref{eq:bzuv}). From the entropy one can derive the pressure by
solving $dp(T)/dT=s(T)$. Another possibility is to compute the free energy from the Einstein-Hilbert action. One has to regularize the action by introducing a cutoff $z=\epsilon$  in the integral over the on shell action Eq.~(\ref{eq:action5D}), c.f. Ref.~\cite{Gursoy:2008za}. The free energy is computed as the difference between the free energy of the BH solution and that of the thermal gas solution. The result is~\cite{Gursoy:2008za,Megias:2010ku}
\begin{equation}
{\cal F} = \frac{1}{\beta} \lim_{\epsilon \to 0} ( {\cal
S}^{\textrm{\tiny BH}}_{\textrm{\tiny reg}}(\epsilon) - {\cal
S}^{\textrm{\tiny TG}}_{\textrm{\tiny reg}}(\epsilon) ) = \frac{\textrm{Vol}(3)}{16\pi G_5} \left( 15 G -
\frac{C_f}{4} \right) \,, \label{eq:Fk}
\end{equation}
where $C_f = 4\pi T b^3(z_h)$ corresponds to an enthalpy contribution
in ${\cal F}$, and~$G = \frac{\pi G_5}{15}
\frac{\beta(\alpha)}{\alpha^2} \left( \langle \textrm{Tr} F^2_{\mu\nu}
\rangle_T - \langle \textrm{Tr} F^2_{\mu\nu} \rangle_0 \right)$ is the
gluon condensate up to normalization factors. This method to compute
${\cal F}$ is more involved giving the fact that the computation of
the gluon condensate $G$ is rather subtle. The finite temperature
solution of the Einstein equations differs from the zero temperature
one at~${\cal O}(z^4)$:
\begin{equation}
\frac{b(z)}{b_0(z)} = 1 + \frac{G}{\ell^3}z^4\left( 1 + \frac{19}{12}\beta_0\alpha_0(z) + {\cal O}(\alpha^2_0) \right) + \cdots
\label{eq:bz4}
\end{equation}
In Eq.~(\ref{eq:bz4}) care has to be taken to keep track of
logarithmic effects which are not usually considered,
c.f. Ref.~\cite{Gursoy:2008za}. These higher order terms in $\alpha_0$
affect appreciably the extraction of $G$ from a comparison of the
numerical solutions of $b$ and $b_0$. Only when at least the order
${\cal O}(\alpha_0)$ is taken into account in Eq.~(\ref{eq:bz4}), one
gets good agreement of the thermodynamic quantities with the numerical
results from the Bekenstein-Hawking entropy formula
Eq.~(\ref{eq:s1text}). We show in Fig.~\ref{fig:freeenergy12} the free
energy obtained by using the Bekenstein-Hawking entropy formula,
Eq.~(\ref{eq:s1text}), and the Einstein-Hilbert action from
Eq.~(\ref{eq:Fk}), using NNLO terms in $\alpha_0$ in
Eq.~(\ref{eq:bz4}) to compute $G$ as a function of~$T$. These plots
correspond to a numerical solution of the equations of motion. See
Ref.~\cite{Megias:2010ku} for details. One recognizes in this figure
the first order phase transition at $T_c=273 \,\textrm{MeV}$ for zero
flavours, which is quite close to lattice simulations. The upper
branch in Fig.~\ref{fig:freeenergy12} represents the small black
holes.

\section{Spatial string tension and Polyakov loop}
\label{sec:spatial_Polyakov}

\begin{figure}[tbp]
\centerline{\includegraphics[width=7.3cm,height=5cm]{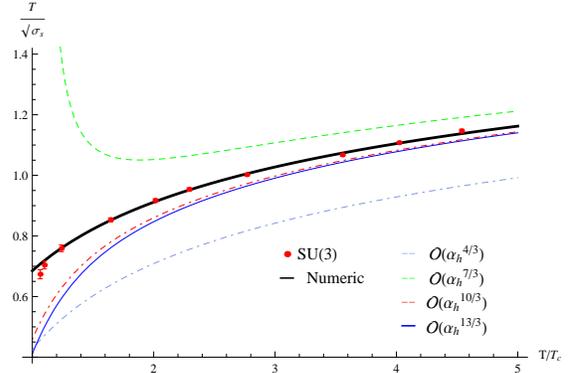}}
\caption{$T/\sqrt{\sigma_s}$ as a function of $T$ (in units of $T_c$). We show as points the lattice data for SU(3) taken from Ref.~\cite{Boyd:1996bx}. The colored curves represent the analytical result from the holographic model, c.f. Eq.~(\ref{eq:sigmaswc}), and the black solid line refers to the full numerical result of Eq.~(\ref{eq:sigmas1}).
}
\label{fig:sigmas}
\end{figure}

The spatial string tension is very useful to test AdS/QCD models. It
is non-vanishing even in the deconfined phase, and it gives useful
information about the non perturbative features of high temperature
QCD.  The computation of the correlation function of rectangular
Wilson loops in the $(x,y)$ plane leads to a potential between quark
and antiquarks which behaves linearly at large distances, i.e.
\begin{equation}
\langle W[{\cal C}] \rangle \stackrel{y\to\infty} \simeq  e^{-y \cdot V(d)} \,, \qquad
V(d) \stackrel{d\to\infty} \simeq \sigma_s \cdot d  \,.
\end{equation}
For details on the computation we refer to e.g. Ref.~\cite{Alanen:2009ej}. The spatial string tension takes the following form:
\begin{equation}
\sigma_s(T) = \frac{1}{2\pi l_s^2} \alpha_h^{4/3} b^2(z_h) \,. \label{eq:sigmas1}
\end{equation}
From Eqs.~(\ref{eq:bzuv}) and (\ref{eq:sigmas1}), the ultraviolet expansion leads to
\begin{eqnarray}
&&\sigma_s(T) = \frac{\ell^2}{2 l_s^2} \pi T^2 \alpha_h^\frac{4}{3} \times \label{eq:sigmaswc} \\
&&\quad \times \Bigg[ 1 - \frac{8}{9}\beta_0\alpha_h
+\frac{2}{81}(25\beta_0^2-2\beta_1)\alpha_h^2 + 
{\cal O}(\alpha_h^3) \Bigg] \,. \nonumber
\end{eqnarray}
We show in Fig.~\ref{fig:sigmas} a plot of $T/\sqrt{\sigma_s}$ as a
function of temperature including several orders in
Eq.~(\ref{eq:sigmaswc}), and the full numerical computation from
Eq.~(\ref{eq:sigmas1}). One can see that the model reproduces very
well the lattice data in the regime $ 1.10 <T/T_c < 4.5$. A good fit
to the lattice data taken from Ref.~\cite{Boyd:1996bx} is obtained by
using $l_s =1.94 \,\textrm{GeV}^{-1}$.

\begin{figure}[tbp]
\centerline{\includegraphics[width=7.3cm,height=5cm]{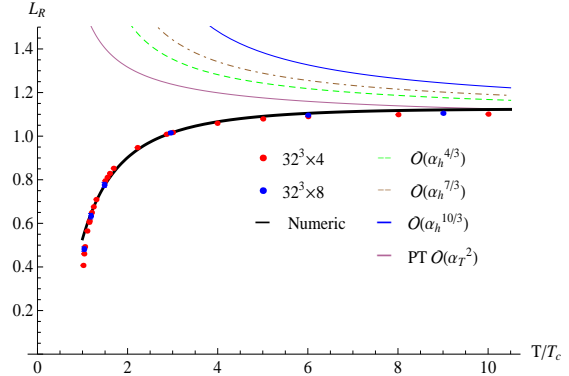}}
\caption{Renormalized Polyakov loop as a function of $T$ (in units of $T_c$). Full (black) line corresponds to the numerical computation of Eqs.~(\ref{eq:Sngreg1})-(\ref{eq:Lreg1}). We show as points lattice data for SU(3) taken from Ref.~\cite{Gupta:2007ax}. The colored curves represent the analytical result from the holographic model, c.f. Eq.~(\ref{eq:L3uvtext}). We also plot the result from pQCD up to ${\cal O}(\alpha_T^2)$ given in Refs.~\cite{Burnier:2009bk,Brambilla:2010xn}.}
\label{fig:PLnf3new2}
\end{figure}

The vacuum expectation value of the Polyakov loop serves as an order
parameter for the deconfinement transition in gluodynamics. The
correlation function of two Polyakov loops taken in the large distance
limit leads to the vacuum expectation value of one single Polyakov
loop squared. This means that the Polyakov loop is related to the free
energy of a single quark~$F_q$ as
\begin{equation}
\langle {\cal P}(\vec{x}) \rangle = e^{-\frac{1}{T}F_q(\vec{x})} \,.
\end{equation}
One can compute the Polyakov loop from the Nambu-Goto (NG) action of a
string hanging down from a quark on the boundary into the
bulk. The NG action reads~\cite{Megias:2010ku,Andreev:2009zk}
\begin{eqnarray}
S^{\textrm{\tiny NG}}_{\textrm{\tiny reg}} &=& \frac{1}{2\pi l_s^2 T}
\bigg[ \int_0^{z_h} dz \; \alpha^{4/3}(z) b^2(z) \nonumber \\
&&\qquad\qquad - \int_0^{z_c} dz \;
\alpha_0^{4/3}(z) b_0^2(z) \bigg] \,, \label{eq:Sngreg1}
\end{eqnarray}
where we have reguralized it by substracting the action of the thermal gas solution up to a cutoff $z_c$. The renormalized Polyakov loop then writes
\begin{equation}
L_R(T) = e^{-S^{\textrm{\tiny NG}}_{\textrm{\tiny reg}} }  \,. \label{eq:Lreg1}
\end{equation}
We show in Fig.~\ref{fig:PLnf3new2} the behavior of $L_R$ as a
function of~$T$ computed numerically from
Eqs.~(\ref{eq:Sngreg1})-(\ref{eq:Lreg1}). We perform a fit to lattice
data by considering the string length $l_s$ and~$z_c$ as free
parameters. The best fit in the regime $T_c< T < 10 T_c$ leads to~$l_s
= 2.36 \, \textrm{GeV}^{-1}$, $z_c = 0.43 \, \textrm{GeV}^{-1}$. Our
approach fits $L_R$ very well without a dimension two condensate,
since a dimension two operator would show up in the UV expansion of
the metric near $z=0$, Eq.~(\ref{eq:bz4}). This does not exclude that
a good fit to the data exists of the form $-2\log L_R \simeq a + b
(T_c/T)^2 $ with $a=-0.23$, $b=1.60$, in accordance with
Ref.~\cite{Megias:2005ve}. See also Ref.~\cite{Andreev:2009zk}.

The UV asymptotic behavior of the Polyakov loop can be computed from Eqs.~(\ref{eq:bzuv}), (\ref{eq:Sngreg1}) and (\ref{eq:Lreg1}). It reads
\begin{equation}
L_R(T) = \exp\left[\frac{\ell^2}{2 l_s^2} \alpha_h^{\frac{4}{3}}
\left( 1 + \frac{4}{9}\beta_0\alpha_h + {\cal O}(\alpha_h^2) \right)\right] \,. \label{eq:L3uvtext}
\end{equation}
We plot in Fig.~\ref{fig:PLnf3new2} also the UV asymptotics given by
Eq.~(\ref{eq:L3uvtext}) and the pQCD result up to ${\cal
  O}(\alpha_T^2)$~\cite{Burnier:2009bk,Brambilla:2010xn}.

\section{Conclusions}

We study the thermodynamics of the 5D dilaton-gravity model by using a
parameterization for the dilaton potential which was quite successful
to reproduce the $Q\bar{Q}$ potential at $T=0$~\cite{Galow:2009kw}. We
compute analytically in the UV and numerically the entropy, free
energy, spatial string tension and Polyakov loop. We demonstrate the
numerical agreement between computations of the free energy from the
Bekenstein-Hawking entropy formula and from the Einstein-Hilbert
action. Both approaches lead to the same result, but the later is much
more sensitive to numerical errors, and an accurate computation is
only possible when one takes care of including logarithmic effects in
an UV expansion of the scale factor at finite $T$.



\begin{theacknowledgments}
  I thank the Humboldt Foundation for their stipend. This work was also supported by the ExtreMe Matter Institute EMMI in the
framework of the Helmholtz Alliance Program of the Helmholtz Association. I thank E. Ruiz Arriola for useful discussions on the Polyakov loop, and H.J.~Pirner for a careful reading of the manuscript.
\end{theacknowledgments}



\bibliographystyle{aipproc}   




\end{document}